\documentclass[
reprint,
 amsmath,amssymb,
 aps,
]{revtex4-2}

\usepackage{graphicx}
\usepackage{lipsum}
\usepackage{comment}
\usepackage{float}
\usepackage{xcolor}
\usepackage{dcolumn}
\usepackage{bm}

\begin{document}


\title{Analytical solution of Brill waves}

\author{Piotr Koc}

\email{piotr@koc.pl}
\affiliation{Independent researcher,\\ ul. Królowej Jadwigi 29/12, 30-209 Kraków, Poland
}

\date{\today}

\begin{abstract}

A class of analytical solutions of axially symmetric  vacuum  initial  data  for  a  self-gravitating  system  has  been  found.  The active region of the constructed gravitational wave is a thin torus around which the solution is conformally flat. For higher  values  of  gravitational wave  amplitude the resulting hypersurface contains apparent horizons.

\end{abstract}

\maketitle


\section*{Introduction} In 1959 Dieter Brill in his  dissertation \cite{Brill} considered the problem  of axially and  time-symmetric, vacuum initial data for the Einstein equations. Although this is the simplest non-trivial case of vacuum constraints of general relativity, no analytical solutions are known so far. 

Over the last half-century  an enormous effort has been made to study  numerical solutions  of initial data for axially symmetric gravitational waves. Due to the lack of analytical solutions, sophisticated numerical methods were developed \cite{Eppley,Gentle, Karkowski, Korobkin,deOliveira,Sorkin,Hilditch,Hilditch2,Garfinkle}. Some properties of this system have also been deducted by advanced indirect  methods \cite{Wheeler,Holz,Murchadha}. The motivation to search for this solution comes from the fact that until now we have not known any non-trivial  analytical vacuum  initial data in general relativity. 

In this paper I construct a class of explicit analytical  solutions of Brill waves. I assume that the initial hypersurface of vacuum self-gravitating system in the moment of time symmetry is axially symmetric, asymptotically flat and regular. 

\section*{Pure gravitational radiation initial value problem}
\label{sec:2}

Momentarily static vacuum constraint equations  of general relativity reduces to:
\begin{equation}
    ^{(3)}R = 0,
\label{R=0}
\end{equation}
where $^{(3)}R$ denotes 3-dimensional scalar curvature  of the initial hypersurface. The axially symmetric line element  in cylindrical coordinates may be written as follows \cite{Brill}:
\begin{equation}
    ds^2 = \Phi^4(\rho,z) \left(  e^{2q(\rho,z)}(d\rho^2 +  dz^2 ) + \rho^2 d\varphi^2 \right).
\label{metryka}
\end{equation}
For the above metric the Hamiltonian constraint (\ref{R=0}) takes the form  of coupled Schr\"{o}dinger-like and  Poisson equations: 
\begin{equation}
  \Delta_3 \Phi + V  \Phi =0,  
\label{Shrodinger}
\end{equation}
\begin{equation}
\Delta_2 q = 4V.  
\label{Poisson}
\end{equation}
The same function $V$ acts respectively as a potential and source in the corresponding above  equations.  ${\Delta}_3$ denotes 3-dimensional flat Laplace operator while ${\Delta}_2$ is 2-dimensional flat laplacian on $\rho z$ plane. Regularity of metric on the axis and asymptotic flatness  implies the boundary conditions for $\Phi$ and $q$:

\begin{equation}
\begin{split}
     q(0,z)=0,\quad &{\frac {\partial q}{\partial \rho}(0,z)}=0, \quad 
    {\frac {\partial q}{\partial z}(\rho, 0)}=0,\\ q=&\mathcal{O}(r^{-2}) \textrm{ as } r\rightarrow\infty,
\label{boundaryq}
\end{split}
\end{equation}

\begin{equation}
\begin{split}
      {\frac {\partial \Phi}{\partial \rho}(0,z)}&=0, \quad   {\frac {\partial \Phi}{\partial z}(\rho ,0)}=0,\\ \Phi\rightarrow  1 &+ \frac{m} {2r} \textrm{ as } r\rightarrow\infty,
\label{boundaryPhi}
\end{split}
\end{equation}
where $r=\sqrt{\rho^2+z^2}$ and $m$ is ADM mass of the system. Moreover the determinant of the metric must have no zeros. Therefore solutions for $\Phi$ have to be everywhere positive $\Phi>0$.

The most common approach to this system of equations is to choose the function $q(\rho,z)$ that meets the  boundary conditions (\ref{boundaryq}) and then solve for $\Phi(\rho,z)$. Such a procedure is effective only through a numerical approach and has been extensively applied in many previous studies \cite{Eppley,Gentle, Karkowski, Korobkin,deOliveira,Sorkin,Hilditch,Garfinkle}. Brill \cite{Brill}, Wheeler \cite{Wheeler}, Holz \textit{et al} \cite{Holz}, Beig and Murchadha \cite{Murchadha} proved many interesting properties of the described system but they have not found
any $q(\rho,z)$ for which (\ref{Shrodinger}) could be solved analytically. 

In this research I propose an alternative  approach. We first select the appropriate function $V$  and then solve equations (\ref{Shrodinger}) and (\ref{Poisson}) for $q$ and $\Phi$. The main problem is how to choose $V$ so that the resulting $q$ would meet the boundary conditions (\ref{boundaryq}). 

The plan of this work is as follows. We will change variables of equations (\ref{Shrodinger}) and (\ref{Poisson}) to toroidal coordinates. Next, I will propose an appropriate function $V$ that will generate $q$ satisfying the boundary conditions (\ref{boundaryq}).
Subsequently, the solution of equations (\ref{Poisson}) and (\ref{Shrodinger}) for $q$ and $\Phi$ will be constructed. Lastly in the resulting analytical solution I will numerically analyze the existence and properties of apparent horizons.

\section*{ Choice of toroidal potential $V$}

Let's transform equations (\ref{Shrodinger}) and (\ref{Poisson}) from the cylindrical to the toroidal coordinates:
\begin{equation}
    \rho={\frac {a\sinh \nu }{\cosh \nu -\cos u }},\qquad z={\frac {a\sin u }{\cosh \nu -\cos u }}
\label{toroidalcoor}
\end{equation}

Parameter $a$ is a major radius of the torus. Azimuth angle $\varphi$ is the same in both coordinate systems. Now equations (\ref{Shrodinger}) and (\ref{Poisson}) take the following form:
\begin{widetext}
\begin{equation}
    \frac{(\cosh \nu-\cos u)^3}{a^2\sinh \nu }
    \left[\frac {\partial}{\partial u} 
    \left({\frac {\sinh \nu }{\cosh \nu -\cos u }}  \frac {\partial}{\partial u} \right)+
    \frac {\partial}{\partial \nu} 
    \left({\frac {\sinh \nu }{\cosh \nu -\cos u }}  \frac {\partial}{\partial \nu} \right)\right]\Phi + V\Phi = 0,
\label{toroidalShoringer}
\end{equation}

\begin{equation}
  \frac {({\cosh \nu -\cos u })^2}{a^2} \left(\frac {\partial^2}{\partial u^2} +\frac {\partial^2}{\partial \nu^2} \right)q=4V.
\label{toroidalPoisson}
\end{equation}
\end{widetext}

Respectively also transform the boundary conditions (\ref{boundaryq}) and (\ref{boundaryPhi}).

Let's choose the  function $V$ such that it vanishes outside the thin torus. So assume that the minor radius of the torus $R_2$ is  negligibly small compared to the large one: $R_2\ll a$. The interior of this torus will be \textit{an active region}  of a constructed gravitational wave. Near  the inner circle of the torus  $e^\nu\gg 1$. So it would be more convenient to specify a new variable $\tau$, which measures the distance from the inner circle of the torus:
\begin{equation}
\tau=2ae^{-\nu}.
\label{tau}
\end{equation}
Additionally, let's assume that  $V$ forms a double-well potential  inside toroidal active region:
\begin{figure}[ht]
\centering
\includegraphics[width=\linewidth]{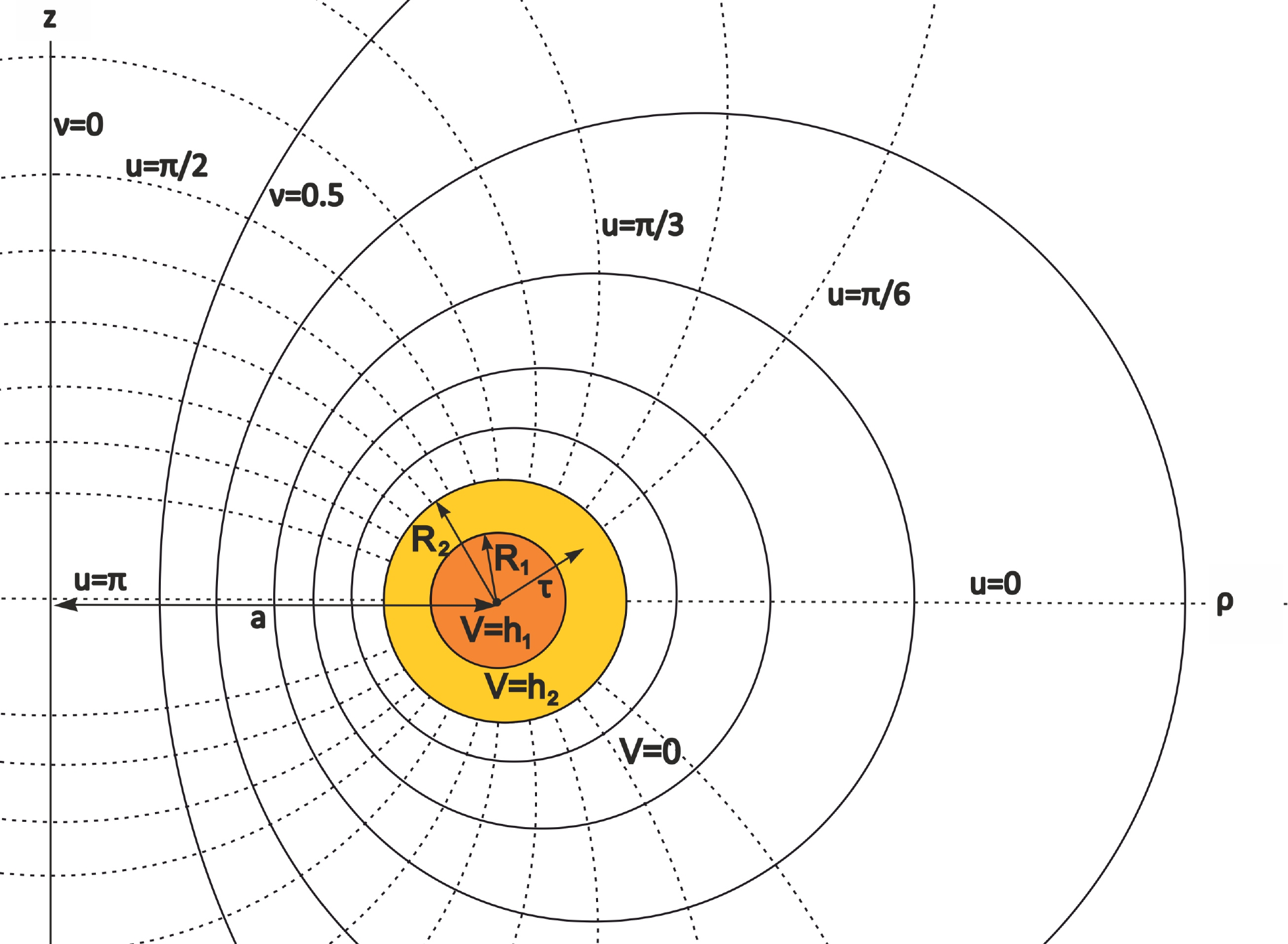}
\caption{Function $V$ in toroidal coordinates $\nu,u$. Dotted circles are surfaces of constant $u$  and solid ones are toruses of constant $\nu$. Potential $V$ vanishes outside of toroidal active region $\tau >R_2$. Inside active region $V$ forms double-well potential with values $V_1$ and $V_2.$}
\label{fig1}
\end{figure}

\begin{equation}
V(\tau,u)=  \begin{cases}   
h_1 &\mbox{for } \tau \leq R_1 \\
h_2 &\mbox{for } R_1<\tau \leq R_2 \\
0 & \mbox{for } \tau > R_2. \end{cases}
\label{Vassume}
\end{equation}

Potential $V$ is depicted schematically in Fig. \ref{fig1}. In the next section, analyzing solution for $q$, I will choose the constants $h_1$ and $h_2$ such that also function $q$ would vanish outside active region $\tau>R_2$.

\section*{Solution for function $q$}
 Wheeler \cite{Wheeler} interpreted function $q(\rho, z)$ as "distribution of gravitational wave amplitude". In constructed solution it will be concentrated only on a thin toroidal active region. Inside our thin torus  $e^\nu\gg 1$. In such a limit equation (\ref{toroidalPoisson}) simplifies to:

\begin{equation}
  \frac {e^{2\nu}}{4a^2} \left(\frac {\partial^2}{\partial u^2} +\frac {\partial^2}{\partial \nu^2} \right)q=4V.
\label{Poissonthin}
\end{equation}
Changing variable to $\tau=2ae^{-\nu}$ we get

\begin{equation}
 \frac{\partial^2 q}{\partial \tau^2} +
 \frac{1}{\tau}  \frac {\partial q}{\partial \tau} +
 \frac{1}{\tau^2} \frac {\partial^2 q}{\partial u^2}=4V.
\label{Poissontau}
\end{equation}
This equation is solved by:

\begin{equation}
q(\tau,u)=  \begin{cases}   
c_1 \ln(\tau) + c_2 + h_1 \tau^2 &\mbox{for } \tau \leq R_1 \\
d_1 \ln(\tau) + d_2 + h_2 \tau^2 &\mbox{for } \tau \in (R_1, R_2) \\
0 & \mbox{for } \tau > R_2. \end{cases}
\label{qintro}
\end{equation}

We have four conditions for the continuity of the function $q$ and its derivative at $R_1$ and $R_2$ and additionally the requirement of regularity at $\tau=0$. This way, we get all constants of integration and an additional requirement that must be met by potential $V$:
\begin{equation}
\begin{split}
& c_1=0,\qquad\qquad   c_2 =2 h_2 R_2^2 \ln\left(\frac{R_2}{R_1}\right), \\[2ex]
& d_1 = - 2 h_2 R_2^2,\quad  d_2 =h_2 R_2^2 (2\ln(R_2)-1),
\label{qconst}
\end{split}
\end{equation}
\begin{equation}
    h_1=h_2\left(1-\frac{R_2^2}{R_1^2}\right).
\label{hconst}
\end{equation}
Thanks to the last condition the influence of positive and negative wells of function $V$  is averaged in such a way that $q$ vanishes outside the active toroidal region and boundary conditions (\ref{boundaryq}) are naturally met. This property that "volume average of potential is zero" have been proved  by Wheeler \cite{Wheeler}.  After returning to the variable $\nu$ and inserting appropriate constants (\ref{qconst})  and (\ref{hconst}) to (\ref{qintro}), we finally get:
\begin{widetext}
\begin{equation}
q(\nu,u)=  2 h_2\cdot\begin{cases}   
R_2^2 \ln\left(\frac{R_2}{R_1}\right)  + 2\left(1-\frac{R_2^2}{R_1^2}\right) a^2e^{-2\nu} &\mbox{for } 2ae^{-\nu} \leq R_1 \\[2ex]
 R_2^2 \left[\ln\left(\frac{R_2}{2a\sqrt{e}}\right)+\nu\right] +  2a^2e^{-2\nu}  &\mbox{for } R_1<2ae^{-\nu} \leq R_2 \\[2ex]
0 & \mbox{for } 2ae^{-\nu} > R_2.\end{cases}
\label{qfinal}
\end{equation}
\end{widetext}
\section*{Solution for function $\Phi$}
 Again we start from the thin torus active region. Inside the torus $e^\nu\gg 1$ so the equation (\ref{toroidalShoringer}) simplifies to

\begin{equation}
    \frac{e^{2\nu}}{4a^2}  \left(\frac {\partial^2}{\partial u^2} +\frac {\partial^2}{\partial \nu^2} \right)\Phi + V\Phi = 0.
\label{Phismall}
\end{equation}
Changing variable to $\tau=2ae^{-\nu}$ we get

\begin{equation}
 \frac{\partial^2 \Phi}{\partial \tau^2} +
 \frac{1}{\tau} \frac{\partial \Phi}{\partial \tau} +
 \frac{1}{\tau^2} \frac {\partial^2 \Phi}{\partial u^2}+ V\Phi=0.
\label{Phitau}
\end{equation}
For our double-well potential (\ref{Vassume}) above equation reduces to Helmholtz equation and may be solved by:

\begin{equation}
\Phi(\tau,u)=  \begin{cases}   
\bar{c}_1J_0(\sqrt{h_1}\tau)+\bar{c}_2Y_0(\sqrt{h_1}\tau) &\mbox{for } \tau \leq R_1 \\[2ex]
\bar{d}_1J_0(\sqrt{h_2}\tau)+\bar{d}_2Y_0(\sqrt{h_2}\tau) &\mbox{for } \tau \in (R_1, R_2)  \end{cases}
\label{Phiinterior}
\end{equation}
where $J_0$ and $Y_0$ are Bessel functions of the first and second kind.

Outside active region $V=0$ so equation (\ref{toroidalShoringer}) reduces to Laplace equation that in toroidal coordinates is separable by standard methods \cite{Loh}. Taking into account the boundary conditions (\ref{boundaryPhi}),  and local polar symmetry of the active region, the solution has to take the form:

\begin{equation}
    \Phi(\nu,u)=1 + \frac{m}{2a\sqrt{2}}\sqrt{{\cosh \nu - \cos u}} \ P_{-\frac{1}{2}}(\cosh \nu)
\label{Phiout1}
\end{equation}
$P_{-\frac{1}{2}}$ is Legendre function with half-integer index that is also known as toroidal harmonic. This particular harmonic $P_{-\frac{1}{2}}$ is related to the elliptic integral of the first kind: $P_{-\frac{1}{2}}(x)=\frac{2}{\pi} K\left(\frac{1-x}{2}\right)$. 
Same as before we have to glue function $\Phi$ smoothly on $R_1$ and $R_2$. Near the active region where $e^\nu\gg 1$  equation (\ref{Phiout1}) simplifies to:
\begin{equation}
 \Phi(\nu,u)= 1+\frac{m }{2 \pi  a}\ln \left(\frac{e^{\nu }}{2}\right)=1+\frac{m}{2\pi a} \ln {\frac{a}{\tau}}
\label{Phinear}
\end{equation}

The regularity of the function $\Phi$ in equation (\ref{Phiinterior})  at  $\tau=0$  implies that $\bar{c}_2=0$. Using the above formula and (\ref{Phiinterior}) we may impose continuity conditions on the function $\Phi$ and its derivative at $R_1$ and $R_2$. Taking into account  also condition (\ref{hconst}), we get the following set of equations:


\begin{widetext}
\begin{equation}
\begin{split}
    &\bar{c}_1J_0\left(\sqrt{h_2\left(1-\frac{R_2^2}{R_1^2}\right)}R_1\right)= \bar{d}_1J_0(\sqrt{h_2}R_1)+\bar{d}_2Y_0(\sqrt{h_2}R_1),\\ 
    &\bar{d}_1J_0(\sqrt{h_2}R_2)+\bar{d}_2Y_0(\sqrt{h_2}R_2) =1+\frac{m}{2\pi a} \ln {\frac{a}{R_2}}, \\  
    &\bar{c}_1\sqrt{h_2\left(1-\frac{R_2^2}{R_1^2}\right)} J_1\left(\sqrt{h_2\left(1-\frac{R_2^2}{R_1^2}\right)} R_1\right)=\bar{d}_1\sqrt{h_2} J_1(\sqrt{h_2} R_1) +\bar{d}_2\sqrt{h_2} Y_1(\sqrt{h_2} R_1),\\  &\bar{d}_1\sqrt{h_2} J_1(\sqrt{h_2} R_2) +\bar{d}_2\sqrt{h_2} Y_1(\sqrt{h_2} R_2)=\frac{m}{2\pi a R_2}.
    \label{Phiconst} 
\end{split}
\end{equation}
\end{widetext}

Explicit solution (for $\bar{c}_1$, $\bar{d}_1$, $\bar{d}_2$ and $m$) of this system of equations leads to obvious but extremely lengthy formulas.  After returning to the  variable $\nu$ finally $\Phi$ takes the following form:
\begin{widetext}
\begin{equation}
\Phi(\nu,u)=  \begin{cases}   
\bar{c}_1J_0\left(2\sqrt{h_2\left(1-\frac{R_2^2}{R_1^2}\right)}\ ae^{-\nu}\right) &\mbox{for } 2ae^{-\nu} \leq R_1 \\[3ex]
\bar{d}_1J_0(2\sqrt{h_2}\ ae^{-\nu})+\bar{d}_2Y_0(2\sqrt{h_2}\ ae^{-\nu}) &\mbox{for } R_1<2ae^{-\nu} \leq R_2\\[3ex]
1 + \frac{m}{2a\sqrt{2}}\sqrt{\cosh \nu - \cos u}\ P_{-\frac{1}{2}}(\cosh \nu)&\mbox{for } 2ae^{-\nu}>R_2
\end{cases}
\label{Phifinal}
\end{equation}
\end{widetext}

Since $h_1$ and $h_2$ are constrained by the equation (\ref{hconst}), they have opposite signs. Therefore in equations (\ref{Phiconst}, \ref{Phifinal}) some of the  Bessel functions have a purely imaginary argument. Nevertheless all the resulting functions all real because Bessel functions of purely imaginary argument reduce to modified Bessel  functions \cite{abramowitz} that are real. 

\section*{Some properties of the solution}

Lastly, let's also transform the line element (\ref{metryka}) to the toroidal coordinates:
\begin{equation}
    ds^2 = a^2\Phi^4(\nu,u) \frac{e^{2q(\nu,u)}(d\nu^2 +  du^2 ) + \sinh^2 \nu d\varphi^2}{(\cosh \nu -\cos u)^2 }. 
\label{metrykatoro}
\end{equation}
Function $q(\nu, u)$ is specified by equation (\ref{qfinal}). Conformal factor $\Phi(\nu, u)$ is given by equation (\ref{Phifinal}) where constants $\bar{c}_1$, $\bar{d}_1$, $\bar{d}_2$ and $m$ are determined by (\ref{Phiconst}). So the final solution depends on four parameters: $a$, $R_1$, $R_2$ and potential well depth $h_2$ which determines the strength of the field in the active thin toroidal region.

\begin{figure}[H]
\centering
\includegraphics[width=1\linewidth]{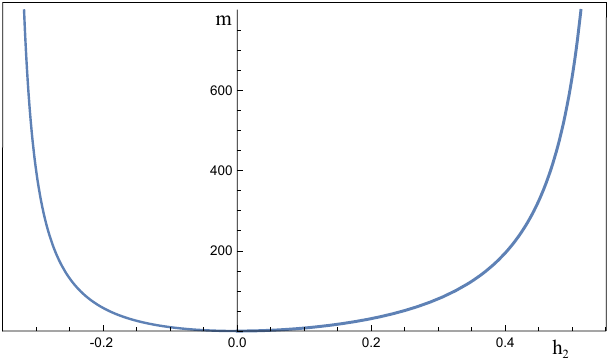}
\caption{Relationship between  ADM mass $m$ and "gravitational  wave  amplitude" parameter $h_2$. For $h_2\approx -0.340$ and  $h_2\approx 0.573$ ADM mass tends to infinity. Higher values of parameter $h_2$ lead to zeros of function $\Phi$ therefore such configurations are unphysical. The other parameters set to: $R_1=1$, $R_2=2$, $a=100$.}
\label{fig2}
\end{figure}

To illustrate the properties of the metric, let's assume that: $R_1=1$, $R_2=2$, $a=100$ and $h_2$ is the only free parameter determining the strength of the gravitational field.  Relationship between   ADM mass $m$ and $h_2$ for these set of parameters is shown in Fig. \ref{fig2}.

Let us also examine the existence and shape of the apparent horizons surfaces $S$. In the momentarily static case the expansion of outgoing orthogonal future directed null geodesics is the divergence of normal to $S$ unit vector $D_i n^i=0$. Outside  thin toroidal active region for our metric (\ref{metrykatoro}) this condition takes the form:

\begin{widetext}
\begin{multline}
   \nu''+ \nu'^3 \left(\frac{4 \Phi_u}{\Phi}-\frac{2 \sin u}{\cosh\nu-\cos u}\right)-\nu'^2 \left(\frac{4 \Phi_\nu }{\Phi}-\frac{2 \sinh \nu}{\cosh\nu-\cos u} +\coth \nu  \right)+\\
   + \nu' \left(\frac{4 \Phi_u }{\Phi}-\frac{2 \sin u}{\cosh\nu-\cos u}\right) 
   -\left(\frac{4 \Phi_\nu  }{\Phi}-\frac{2 \sinh \nu}{\cosh\nu-\cos u}  +\coth \nu\right)  =0
\label{apparenthorison}
\end{multline}
\end{widetext}

\begin{figure}[ht]
\centering
\includegraphics[width=1\linewidth]{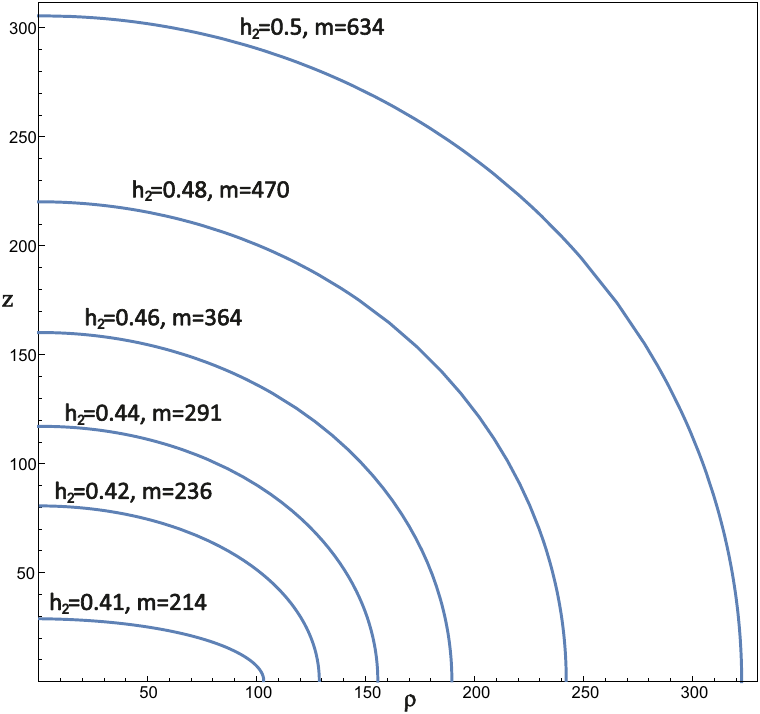}
\caption{Shapes of external apparent horizons for several value of parameter $h_2$ and corresponding ADM mass $m$ of the system. The other parameters are the same as in FIG. \ref{fig2} }
\label{fig3}
\end{figure}

Now using the solution (\ref{Phifinal}) we may solve the above equation numerically. A several examples of external apparent horizons are depicted in  Fig. \ref{fig3}. The mass of the system increases with $h_2$ and the corresponding apparent horizons are further from the center of the system and become more spherical, which is in accordance with our physical intuition. For non-vacuum systems analogical study of trapped surfaces in toroidal geometries  has also been conducted recently by Karkowski \textit{et al} \cite{Karkowski2}.

\section*{Summary}
An analytical solution of Brill waves has been found. Currently, this is the only one known (nontrivial) solution of vacuum initial data in general relativity. In this construction, gravitational wave amplitude $q$ is concentrated in the thin toroidal region. Coefficients of the metric are given in terms of elementary functions and the elliptic integral.   Numerical analysis of the obtained analytical solution shows the existence of apparent horizons for some higher values of gravitational wave amplitude $q$. Further studies might include the analysis of the evolution of these initial data.

\par
\section*{Acknowledgements}
The author would like to thank Anna Klecha, Wojciech Grygiel and Tadeusz Pałasz for discussions and reading the manuscript.

\bibliography{ref}

\end{document}